\documentclass[aps,prb,twocolumn,showpacs,preprintnumbers,amsmath,amssymb,superscriptaddress]{revtex4-1}

\usepackage{graphicx}
\usepackage{dcolumn}
\usepackage{bm}
\newcommand{\nc}{\newcommand}
\nc{\be}{\begin{equation}}
\nc{\ee}{\end{equation}}
\nc{\bea}{\begin{eqnarray}}
\nc{\eea}{\end{eqnarray}}
\nc{\bean}{\begin{eqnarray*}}
\nc{\eean}{\end{eqnarray*}}
\nc{\mb}{\mbox}
\nc{\rnc}{\renewcommand}
\nc{\vk}{\mb{\bf k}}
\nc{\vp}{\mb{\bf p}}
\nc{\vn}{\mb{\bf n}}
\nc{\vq}{\mb{\bf q}}
\nc{\rr}{\mb{\bf r}}
\nc{\vz}{\hat {\mb{\bf z}}}
\nc{\vj}{\mb{\boldmath$j$}}
\nc{\vg}{\mb{\boldmath$g$}}
\nc{\x}{\mb{\boldmath$x$}}
\nc{\A}{\mb{\boldmath$A$}}
\nc{\va}{\mb{\boldmath$a$}}
\nc{\vs}{\mb{\boldmath$\sigma$}}
\nc{\vpi}{\mb{\boldmath$\pi$}}
\nc{\nab}{\nabla}
\nc{\X}{\sf x}

\begin{document}

\title{Thermopower of Quantum Hall States in Corbino Geometry as a Measure of Quasiparticle Entropy}

\author{Yafis Barlas}
\affiliation{Department of Physics and Astronomy, University of California,
Riverside, CA 92521}
\author{Kun Yang}
\affiliation{National High Magnetic Field Laboratory and Department of Physics, Florida State
University, FL 32306, USA}

\begin{abstract}
Using the Onsager relation between electric and heat transport coefficients, and considering the very different roles played by the quantum Hall condensate and quasiparticles in transport, we argue that near the center of a quantum Hall plateau thermopower in a Corbino geometry
measures {\it ``entropy per quasiparticle per quasiparticle charge"}. This relation indicates that thermopower measurement in a Corbino setup is a 
more direct measure of quasiparticle entropy than in a Hall bar. Treating disorder within the self-consistent Born approximation, we show through an explicit microscopic calculation that this relation holds on an integer quantum Hall plateau at low temperatures. Applying this to non-Abelian quantum Hall states, we argue that Corbino thermopower at sufficiently low temperature becomes temperature-independent, and measures the quantum dimension of non-Abelian quasiparticles that determines the topological entropy they carry.
\end{abstract}


\maketitle


\section{Introduction}
\begin{figure}[t]
\begin{center}
\includegraphics[width=2.0in,height=1.8in]{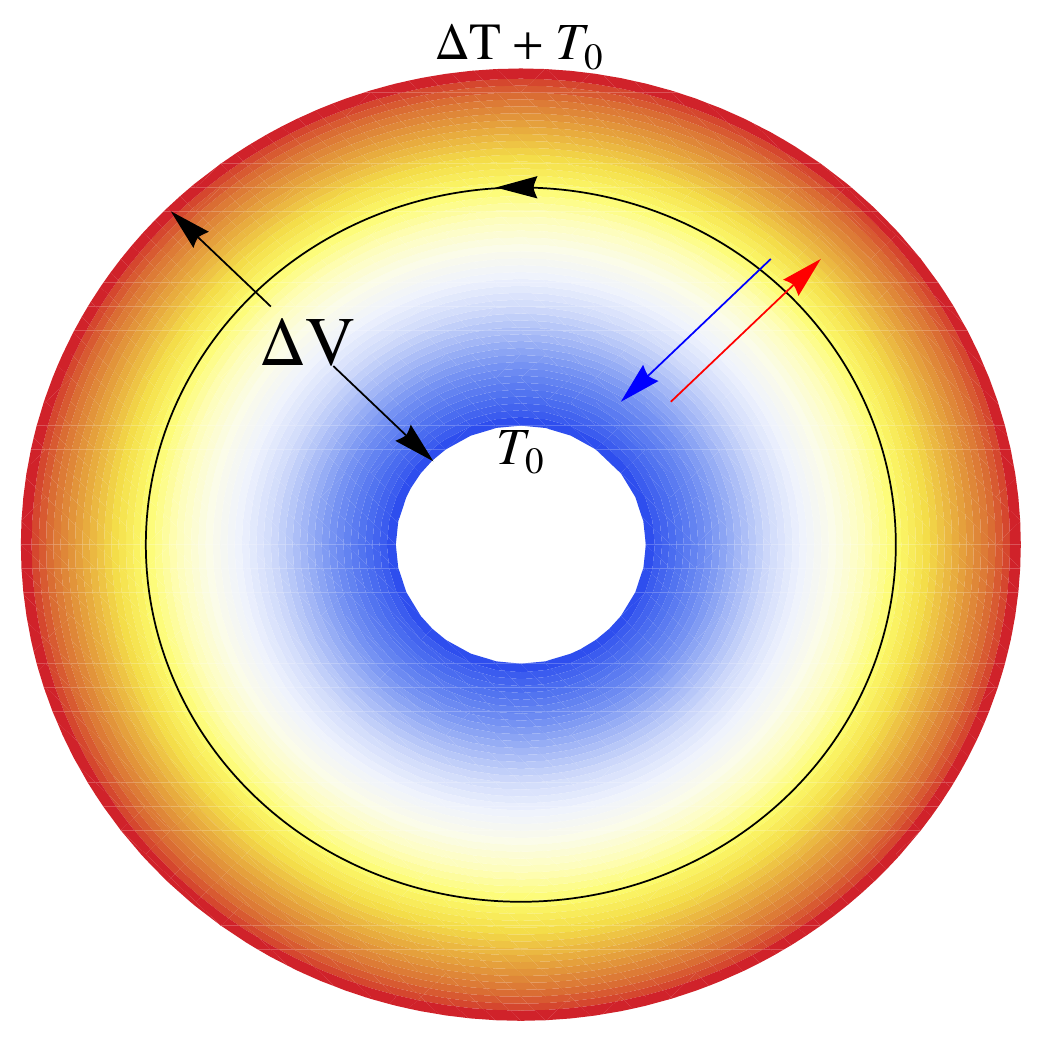}
\caption{(Color online) Schematic representation of thermopower measurement in a Corbino disk geometry with disorder. A thermal gradient applied in the radial direction leads to a corresponding voltage difference, with no net current flowing along the same direction; there is {\em no} constraint on the circular current flowing along the angular direction. Since the condensate current (represented by the black line, including edge current) is always in the Hall channel and thus flowing along the angular direction, it plays {\em no} role in the current balance along the radial direction. As a result the zero radial current condition must be satisfied by balancing quasiparticle current responses to both thermal and voltage gradients, and thus dictated by quasiparticle properties {\em only}. Since the net quasiparticle current is zero, disorder plays a much lesser role here than in Hall geometry.}
\label{figtwo}
\end{center}
\end{figure}

Candidate fractional quantum Hall (FQH) states which exhibit non-Abelian quasiparticle excitations have been predicted to appear in the second Landau level (LL) for certain filling fractions.~\cite{mooreread,readrezayi}
Recent proposals for performing intrinsic fault-tolerant quantum computation using non-Abelian anyons~\cite{kitaev,nayak} have revived interest at these filling factors, and in particular on the nature of their quasiparticles excitations. At present the most promising candidate for non-Abelian FQH state (based on numerical studies~\cite{MRnumerics}) seems to be $\nu = 5/2$, which is thought to be described by the Moore-Read state\cite{mooreread} or its particle-hole conjugate.\cite{lee07} Experimentally, the quasiparticle charge has been measured via tunneling between opposite edge states across quantum point contacts~\cite{heiblum,willet1} as well as local charge measurement,\cite{Venkatachalam} and found to be consistent with theoretical predictions expected for non-Abelian quasiparticles. In particular the observation of an "even-odd" effect~\cite{bonderson} alternating between $e/4$ and $e/2$ quasiparticles,~\cite{willet2} as well as the recently reported phase slips\cite{kang} in the interference pattern are suggestive of the non-Abelian nature of the quasiparticle excitations. However these experiments need to be reconciled with each other, which would require detailed understanding of all aspects of the experiments. Such an understanding requires careful analysis of possible complications due to edge reconstruction,~\cite{kunrezayi} nonequilibrium edge distributions and coupling of the edge state to bulk quasiparticles.\cite{edgebulkcoupling} \\
\indent
In light of possible complications associated with the edge, alternative approaches using {\em bulk} measurements to directly probe the non-Abelian nature of the quasiparticles have been proposed.~\cite{yanghalperin,cooperstern,gervaisyang}
The basic idea behind them is the observation that in the presence of non-Abelian quasiparticles, the system has a ground state degeneracy $\Gamma$ which grows exponentially with the number of quasiparticles $N_{q}$: $\Gamma \sim d^{N_{q}} $ where $d>1$ is the quantum dimension of the non-Abelian quasiparticle. This leads to a {\em temperature independent} entropy (except below an energy scale which is related to the coupling between quasiparticles and decays exponentially with their separation) due to this degeneracy\cite{yangpreprint}:
\begin{equation}
S_{D} = k_{\beta} \log \Gamma  = k_{\beta} N_{q} \log d + O(1).
\end{equation}
This non-Abelian or {\em topological} entropy due to ground state degeneracy associated with the presence of non-Abelian quasiparticles, is an important part of the total entropy:
\begin{equation}
S_{tot} = S_{D}+ S_{n}(T),
\end{equation}
where
$S_{n}(T)$ is the temperature dependent entropy due to normal sources. At sufficiently low temperatures $S_{n}(T)$ approaches zero and $S_D$ dominates $S_{tot}$, thereby allowing experimentalists to measure the topological entropy using probes sensitive to entropy. In particular, it was pointed out in Ref. \onlinecite{yanghalperin} that thermopower is one of the possible probes, because it measures entropy per mobile electron under suitable conditions. In fact, the topological entropy might already be a significant contributor to thermopower measured near 5/2 in a recent experiment.~\cite{caltechexp} 
We note thermopower had been measured in the quantum Hall (QH) regime before that.\cite{obloh,Zalinge}
 \\
\begin{figure}[th]
\begin{center}
\includegraphics[width=2.4in,height=2.0in]{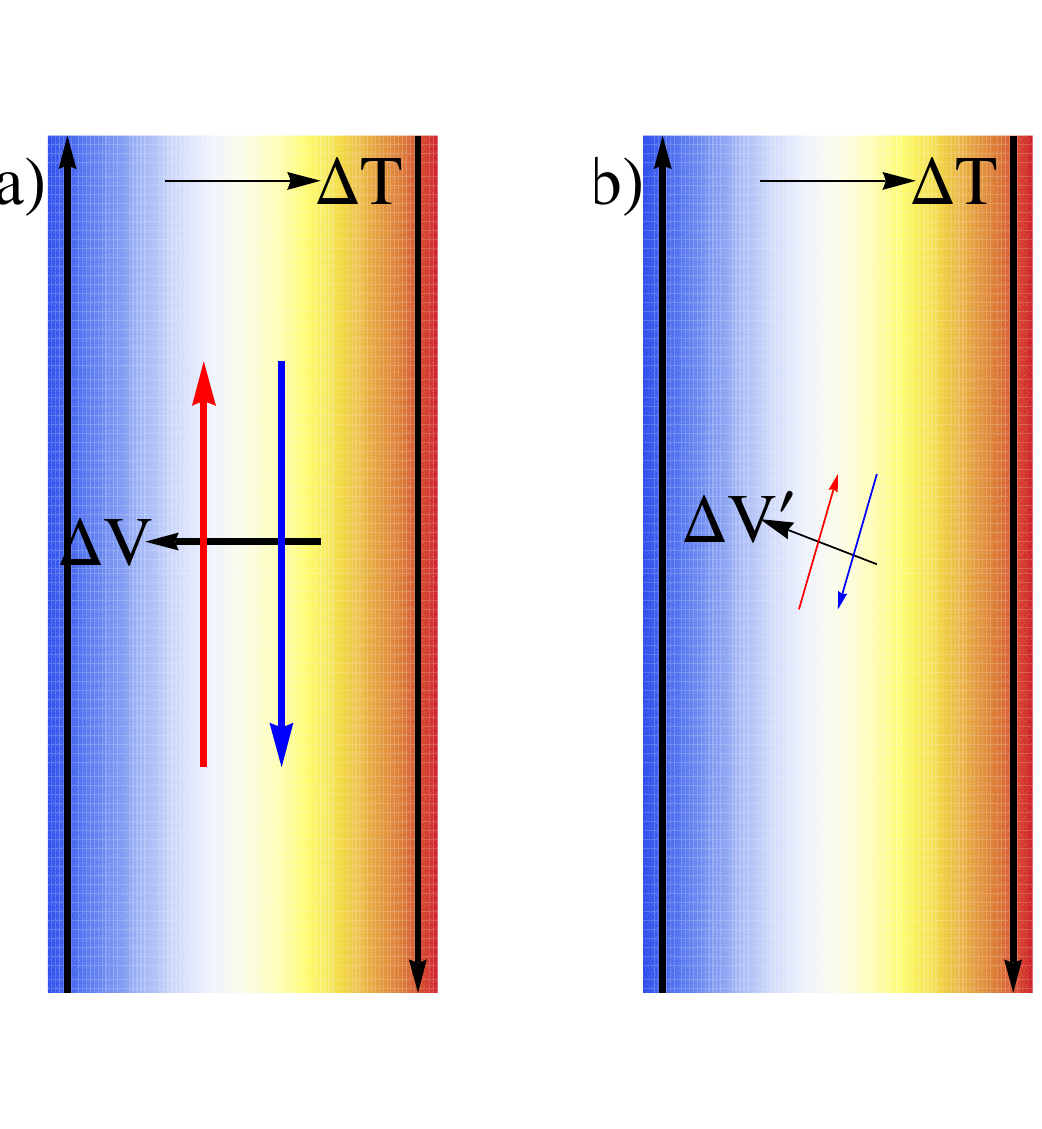}
\caption{(Color online) Schematic representation of current responses to thermal and voltage gradients in an infinite Hall bar setup in (a) clean and (b) disordered limits. The thermal response is due to entropy carried by quasiparticles, while the voltage response is dominated by the quantum Hall condensate. In (a) clean limit both current responses are in transverse (Hall) channel. A thermal gradient applied from the left to right leads to a corresponding voltage difference along the same direction. The current induced by the thermal gradient (red arrow), is canceled exactly by the voltage-induced current dominated by condensate flow (blue arrow), resulting in zero net current flowing through the sample. The equilibrium edge current is represented by the thick black arrows. In the disordered limit (b), thermal gradient induced current is strongly suppressed by pinning or localization of quasiparticles, leading to a corresponding suppression in the compensating condensate current induced by voltage gradient. As a result the voltage response is also suppressed. In the presence of disorder the thermal gradient induced current includes both longitudinal and Hall components, and the voltage gradient is no longer parallel to thermal gradient.}
\label{figone}
\end{center}
\end{figure}
\indent
A QH liquid can be viewed as a macroscopic condensate which has a gap for charged excitations; in the case of a FQH liquid these quasiparticle excitations carry fractional charge and possibly non-Abelian statistics. The QH condensate does not carry any entropy; hence all the entropy must be carried by the quasiparticles. In the presence of a thermal gradient the thermal response (which is proportional to the entropy) can only come from the entropy carried by the quasiparticles; this can be viewed as response to an ``entropical force". In a thermopower measurement the application of a temperature gradient ${\bf \nabla}T $ leads to an electric potential gradient ${\bf \nabla} V$, as no current is allowed to flow through the system. The thermal response must be canceled by the electric response to voltage gradient so the net current is zero. In the Hall bar geometry where the thermopower measurement has been considered earlier,\cite{yanghalperin} the electric response is dominated by the QH condensate. As a result the thermal and electric responses come from very different sources. This difference may lead to substantial {\em suppression} of the thermopower signal when disorder is present, as disorder can have a strong effect on the quasiparticles, potentially leading to their localization and thus suppressing their response to a thermal gradient. On the other hand disorder has virtually no effect on the condensate; a very weak voltage gradient can induce sufficient current response to compensate for the thermal gradient. For this reason the quasiparticle entropy is detectable in Hall bar thermopower measurement only when they are mobile, which requires somewhat elevated temperatures\cite{yanghalperin} (see Fig. \ref{figone} for an illustration of this point).\\
\indent
The purpose of this paper is to point out that thermopower in a Corbino setup\cite{Zalinge} (see Fig. \ref{figtwo}) is a more direct and thus much better measure of quasiparticle entropy. This is because in the Corbino geometry the zero-current condition that results from a cancelation of the thermal and voltage responses only applies to the {\em longitudinal} current (or current along the radial direction); there is {\em no} constraint on the Hall channel current, which simply runs around the annulus. The crucial point here is that the {\em longitudinal} current (from both thermal and voltage gradients) comes from quasiparticles {\em only}, with no condensate or edge state contribution. As a result disorder affects the thermal and electric response {\em on equal footing}, therefore it does {\em not} suppress thermopower even in the presence of localization effects (because we are always working at non-zero temperature, quasiparticles can still hop even if they are localized at $T=0$). Another way to understand this heuristically is that since the quasiparticle current is zero, there is no pinning force on them, thus the electric and entropical forces must be balanced, even in the presence of disorder.\\
\indent
The difference between thermopower in Hall bar and Corbino geometries is best illustrated by analyzing the transport equations in the presence of an electric field and a temperature gradient:
\begin{eqnarray}
\label{transpeq1}
{\bf j} = L^{11} {\bf \nabla} \phi  - L^{12} \frac{{\bf \nabla} T}{T}, \\
\label{transpeq2}
{\bf j}_{Q} = L^{21} {\bf \nabla} \phi  - L^{22} \frac{{\bf \nabla} T}{T},
\end{eqnarray}
where ${\bf j}$ and ${\bf j}_{Q}$ are the respective charge and heat currents, ${\bf \nabla} \phi $ is the applied electric field, $T$ is the temperature and $L^{\alpha \beta} $ are the response coefficients which are tensor quantities in the presence of a magnetic field. The thermopower in a Hall bar geometry, which is given by the zero-current condition in both the transverse and longitudinal directions (${\bf j} =0$), is also a tensor quantity:
\begin{equation}
Q_{H}= \frac{1}{T} L^{12} (L^{11})^{-1};
\end{equation}
it relates ${\bf \nabla} \phi$ with ${\bf \nabla} T$ through
\begin{equation}
{\bf \nabla} \phi= Q_{H} {\bf \nabla} T.
\end{equation}
On the other hand in a Corbino geometry setup a temperature gradient applied in the radial direction leads to a voltage gradient in the radial direction only.
The thermopower which is defined by the zero-current condition in the radial direction is then given as
\begin{equation}
\label{Cth}
Q_{C} = \frac{1}{T}\frac{L^{12}_{rr}}{L^{11}_{rr}},
\end{equation}
which reduces to a number that depends on the longitudinal components of the electric ($L^{11}_{rr}$) and thermal ($L^{12}_{rr}$) responses {\em only}. These longitudinal components are dominated by quasiparticles, and are equally affected by the presence of disorder. The condensate current, which {\it only} flows in the Hall channel is in the angular direction and orthogonal to the radial component, therefore does not contribute to the thermopower measurement.\\
\indent
The central result of this paper is that the Corbino thermopower is (approximately) equal to {\it``entropy per quasiparticle per quasiparticle charge"}:
\begin{equation}
Q_{C}  \approx \frac{ S_{tot}}{e^{\star} N_{q}},
\label{eq:centralresult}
\end{equation}
where $e^{\star}$ is the quasiparticle charge, which is {\em opposite} for electron and hole like quasiparticles.
Eq. (\ref{eq:centralresult}) can be understood heuristically as the consequence of balancing the electric force due to electric field with the ``entropical force" due to the thermal gradient experienced by the quasiparticle:
\begin{equation}
e^{\star}\nabla\phi  = \frac{ S_{tot}}{ N_{q}}\nabla T.
\end{equation}
This is valid when the quasiparticle density is sufficiently low. Since the quasiparticles are {\em not} flowing, the above is expected to be valid even in the presence of disorder, and when the quasiparticles are localized at $T=0$.\\
\indent
In the following sections we explore these ideas in greater detail and use various arguments to establish Eq. (\ref{eq:centralresult}). By applying Onsager relations in the analysis of  transport equations (\ref{transpeq1}) and (\ref{transpeq2}) for the Corbino geometry, we show in section II that in the dilute quasiparticle regime Eq. (\ref{eq:centralresult}) is valid. Since the analysis is quite general and only based on the behavior of the quasiparticle contribution to the transport coefficients it should apply equally for the integer and fractional QH effects. In section III we establish Eq. (\ref{eq:centralresult}) for a specific model of non-interacting electrons in the integer QH regime and presence of disorder through microscopic calculation. In section IV we explore the possibility of using Eq. (\ref{eq:centralresult}) to probe the topological entropy carried by the non-Abelian quasiparticles, and measure their quantum dimension $d$.

\section{Corbino Thermopower}
We consider a QH plateau at the total filling factor $\nu$ in a Corbino geometry setup. At zero temperature for a sufficiently clean and uniform sample, quasiparticles are absent at the center of the QH plateau and the system only consists of the QH condensate. As the magnetic field $B$ is decreased (increased) from its value at the center of the QH plateau $B_{0}$, quasielectrons (quasiholes) are introduced even at zero temperature, whose numbers grow linearly with the deviation of the magnetic field
$\Delta B = B- B_0$:
\begin{equation}
N_{q} = \bigg{|}\bigg(\frac{e}{e^{\star}}\bigg) \frac{\Delta B}{B_{0}}\bigg{|}N_{e},
\end{equation}
where $N_{e}$ is the number of electrons. 
Additional quasiparticles or quasiholes are thermally activated at finite temperature; for sufficiently low temperature one type dominates the other (depending on which side of the plateau one is at). In the following we assume that this is the case and consider the contribution to transport by either quasiparticles or quasiholes {\em only}. \\
\indent
We can split the total charge current ${\bf j}= {\bf j}^{q} + {\bf j}^{cond}$, into a quasiparticle current ${ \bf j}^{q}$ and condensate current ${\bf j}^{cond}$. The latter only flows in the Hall channel and gives rise to quantized Hall transport without dissipation. If the temperature of the inner and outer radii in a Corbino thermopower measurement is fixed so that there is no thermal gradient, then under the application of a radial electric gradient the heat current ${\bf j}_{Q}$, which is carried by quasiparticles only, will be proportional to the quasiparticle charge current.
We can thus write
\begin{equation}
\label{Peltierheat}
{\bf j}_{Q} = \Pi {\bf j}^{q}  = \Pi (e^{\star}{\bf j}^{q}_{n}),
\end{equation}
where ${\bf j}^{q}_{n}$ denotes the quasiparticle number current. The above equation (\ref{Peltierheat}) is similar to the definition of Peltier heat which normally relates the heat current to total current. In our case $\Pi$ relates the heat current to quasiparticle current {\it only} and hence corresponds to the heat carried by each quasiparticle. To proceed further we subtract the QH condensate contribution from the response coefficient of the electric field. As already mentioned in the previous section, only the Hall channel has contributions from both the condensate and quasiparticles, therefore we can separate out the quasiparticle contribution to the Hall conductivity $\tilde{L}^{11}_{xy}$ via,
\begin{equation}
\tilde{L}^{11}_{r \theta} = L^{11}_{r \theta} - \frac{\nu e^2}{h},
\end{equation}
whereas the longitudinal electrical conductivity $L^{11}_{rr} = \tilde{L}^{11}_{rr} $, along with the longitudinal and off-diagonal thermal responses $L^{12}_{rr} = \tilde{L}^{12}_{rr}$ and $L^{12}_{r \theta} = \tilde{L}^{12}_{r \theta}$, are completely dominated by the quasiparticles. In terms of the quasiparticle transport coefficients, $\Pi$ can be expressed as
\begin{equation}
\Pi = \tilde{L}^{21} (\tilde{L}^{11})^{-1}.
\end{equation}
\indent
From the definition (\ref{Peltierheat}) it is clear that $\Pi $ corresponds to the heat carried by each quasiparticle divided by its charge. While in principle it is a tensor, we expect it to be very close to (if not exactly) a pure number (or scalar), since ${\bf j}_Q$ and ${\bf j}^q$ should be parallel to each other on physical ground. We thus focus on its diagonal component $\Pi_{rr}$ in the following. Using the fact $\Pi ({\bf B}) = \Pi (-{\bf B})$ and by virtue of the Onsager relations,~\cite{Onsager} which can be expressed for the quasiparticle contribution to the transport coefficients,
\begin{equation}
\tilde{L}^{\alpha \beta }_{a b} ({\bf B}) = \tilde{L}^{\beta \alpha}_{ b a} (-{\bf B}),
\end{equation}
the diagonal component of the $\Pi$ in the Corbino geometry setup can be expressed as
\begin{equation}
\Pi_{rr} = \frac{\tilde{L}^{11}_{r \theta} \tilde{L}^{21}_{\theta r} + \tilde{L}^{11}_{rr} \tilde{L}^{21}_{rr}}{(\tilde{L}^{11}_{r \theta})^2+(\tilde{L}^{11}_{rr})^2}
= \frac{\tilde{L}^{11}_{r \theta} \tilde{L}^{12}_{\theta r} + \tilde{L}^{11}_{rr} \tilde{L}^{12}_{rr}}{(\tilde{L}^{11}_{r \theta})^2+(\tilde{L}^{11}_{rr})^2}.
\label{eq:Pirr}
\end{equation}
Close to the center of the QH plateau the quasiparticle contributions to the transport coefficients are expected to behave as those of a Hall insulator~\cite{Hallinsulator}: $\tilde{L}^{11}_{rr} \to 0 $ and $\tilde{L}^{11}_{r \theta} \propto (\tilde{L}^{11}_{rr})^2$, which leads to
\begin{equation}
\tilde{L}^{11}_{r \theta} \ll \tilde{L}^{11}_{rr}.
\label{eq:hallinsulatorbehavior}
\end{equation}
Therefore, in the dilute quasiparticle regime or close to the center of the QH plateau the off-diagonal component vanishes much faster than the longitudinal component, we can thus neglect terms involving $\tilde{L}^{11}_{r \theta}$ in Eq. (\ref{eq:Pirr}), which reduces to
\begin{equation}
\Pi_{rr} = \frac{\tilde{L}^{12}_{rr}}{\tilde{L}^{11}_{rr}} = \frac{L^{12}_{rr}}{L^{11}_{rr}} = TQ_{C}.
\end{equation}
The above relation gives an interpretation of thermopower in the Corbino geometry
\begin{equation}
Q_{C} = \frac{\mbox{Heat per quasiparticle}}{e^{\star} T} = \frac{ S_{tot}}{e^{\star} N_{q}},
\end{equation}
which is the {\it "entropy per quasiparticle per quasiparticle charge"}, as advertised in Eq. (\ref{eq:centralresult}). Notice that in the above derivation the condition for a Hall insulator is not essential; to satisfy Eq. (\ref{eq:centralresult}) we only require that close to the center of the QH plateau the quasiparticle contribution to off-diagonal conductivity is much smaller than the diagonal conductivity. This condition on the conductivity (Eq. \ref{eq:hallinsulatorbehavior})can be empirically justified for the case of QH plateaus by examining the experimentally observed relation between the Hall and longitudinal resistivity,\cite{emprel,SimonHalperin}
\begin{equation}
R_{rr} = \alpha_{r} B \frac{dR_{r \theta}}{dB} = \alpha_{r} B \frac{d\Delta R_{r \theta}}{dB}\\
\end{equation}
where $\alpha_{r}$ is an order 1 constant independent of the magnetic field $B$, and $\Delta R_{r \theta}$ is the deviation of Hall resistivity from the quantized plateau value (due to quasiparticle contribution).
Since $\frac{d\Delta R_{r \theta}}{dB}\sim \frac{\Delta R_{r \theta}}{\Delta B}$, we have
\begin{equation}
R_{rr}\sim \frac{B_{0}}{\Delta B} \alpha_r \Delta R_{r \theta} \gg \Delta R_{r \theta}.
\end{equation}
Also due to the fact that on the plateau the quantized Hall conductance due to QH condensate dominates the quasiparticle contributions to the conductivity tensor, the latter is {\em proportional to} the quasiparticle contributions to the resistivity tensor, justifying Eq. (\ref{eq:hallinsulatorbehavior}).

In the next section we lend further support to Eq. (\ref{eq:centralresult}) by performing a microscopic calculation for Corbino thermopower treating disorder in the self consistent Born approximation (SCBA) on an integer QH plateau.
\newline
\section{Corbino Thermopower: IQHE}
In section III A. we calculate the thermopower on an integer quantum plateau in a Corbino geometry with disorder treated within SBCA. In section III B. show that for low temperatures the Corbino thermpower scales like the {\it ``entropy per quasiparticle per quasiparticle charge"}.

\subsection{Corbino Thermopower on an IQH plateau}
\begin{figure}[t]
\begin{center}
\includegraphics[width=3.0in,height=2.6in]{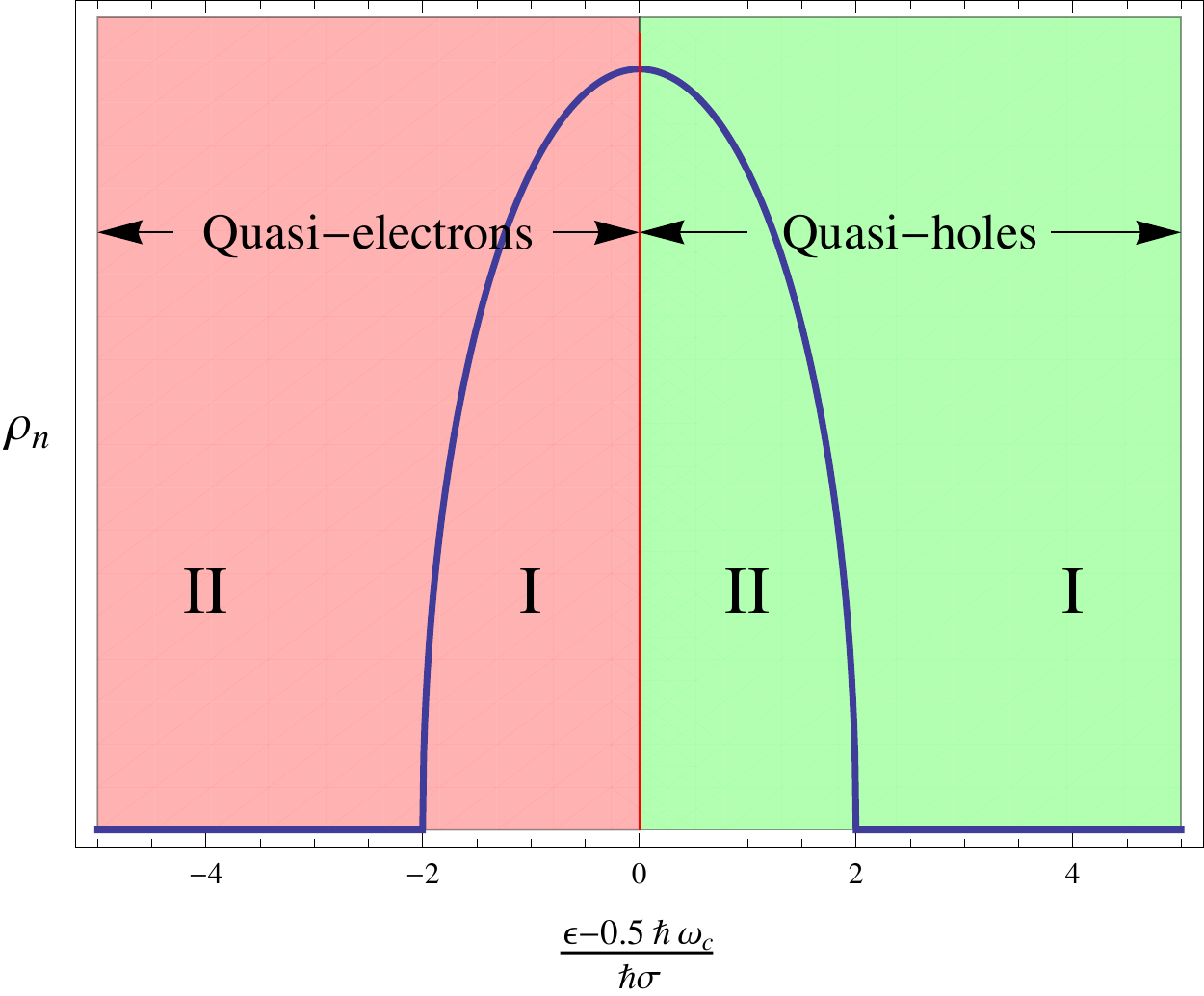}
\caption{(Color on-line) Disorder broadened density of states for the isolated lowest Landau level within self-consistent Born approximation. The red (vertical) line represents the position of the lowest Landau level energy, which is the transition point between two integer quantum Hall phases, and separates the "quasi-electrons" and "quasi-holes" regions. Regions I and II are defined and discussed later in the paper (see text for details).}
\label{figthree}
\end{center}
\end{figure}

\begin{figure}[t]
\begin{center}
\includegraphics[width=3.2in,height=2.6in]{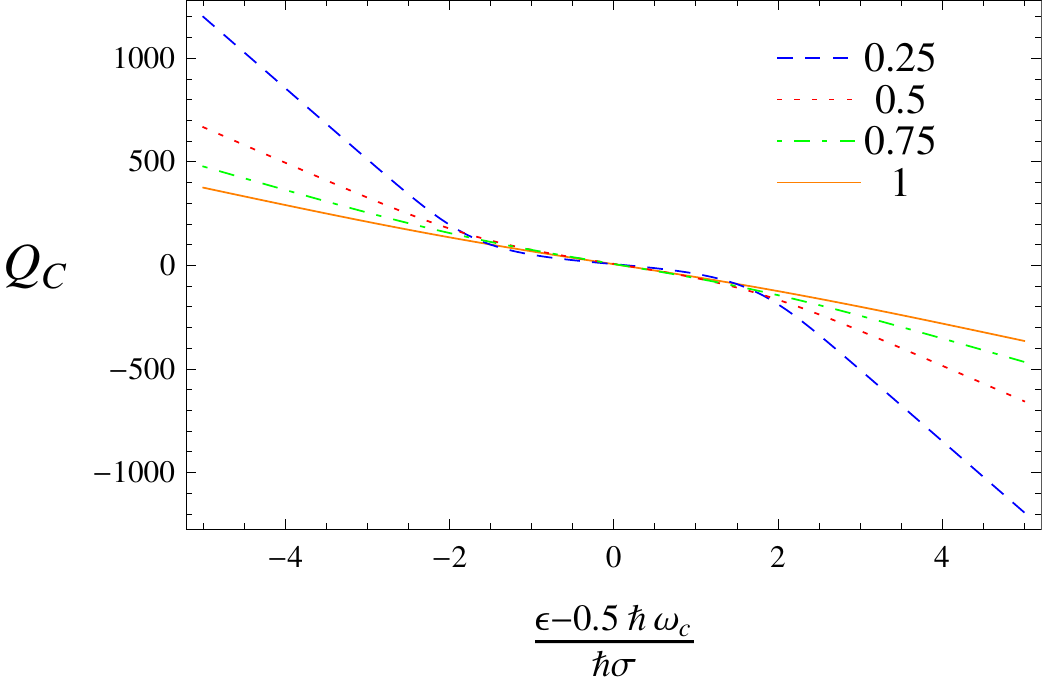}
\caption{(Color on-line) Corbino thermopower measured in units of $\mu V/K$ for the isolated lowest Landau level plotted as a function of $(\epsilon - 0.5 \hbar\omega_{c})/(\hbar\sigma)$, for different values of $\hbar \sigma /k_{\beta} T =0.25,0.5,0.75$ and $1$.}
\label{figfour}
\end{center}
\end{figure}

\begin{figure}[t]
\begin{center}
\includegraphics[width=3.2in,height=2.6in]{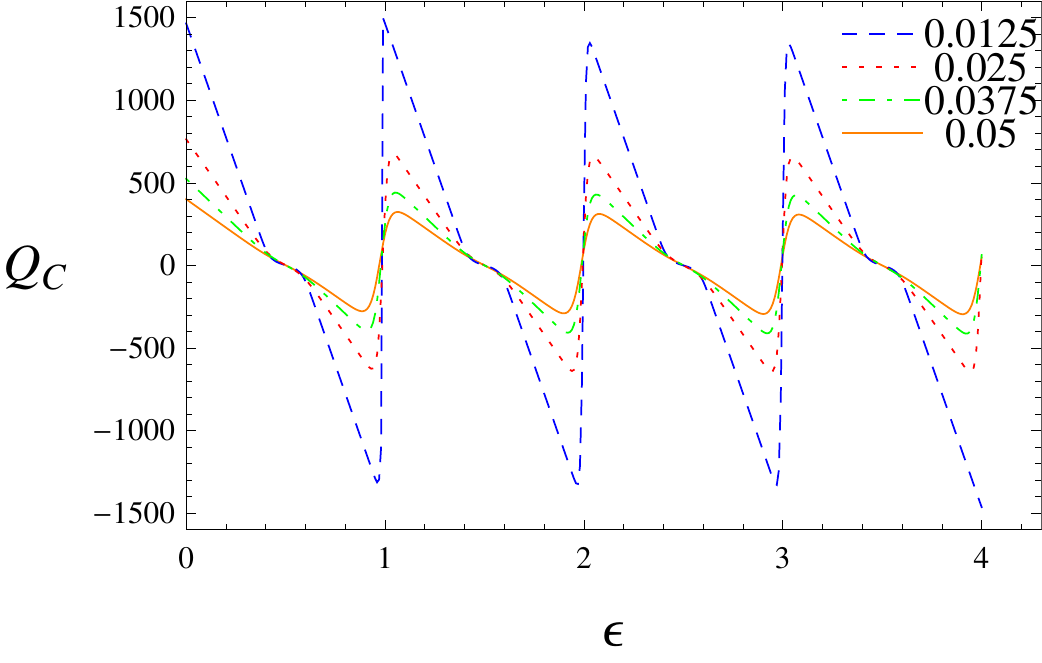}
\caption{(Color on-line) Corbino thermopower calculated for fixed disorder $\hbar \sigma =0.05$ in units of $\hbar \omega_{c}$ at different temperatures $k_{\beta} T = 0.0125,0.025,0.375,$ and $0.05$. The thermopower $Q_{C}$ is measured in $\mu$ V/K and the chemical potential $ \epsilon $ is in units of $\hbar \omega_{c}$.}
\label{figfive}
\end{center}
\end{figure}

The electric field response $L^{11}_{rr}$ can be calculated using the Kubo formula.
We then use the relation~\cite{girvinjonsonprb}
\begin{equation}
\label{ecthrel}
L^{12}_{ij}(T, \mu) = \int_{-\infty}^{\infty}  d \epsilon \frac{\epsilon - \mu}{e}\bigg( -\frac{\partial n_{F}}{\partial \epsilon} \bigg) L^{11}_{ij} (T=0,\epsilon),
\end{equation}
to calculate the thermal response coefficient $L_{rr}^{12}$. In Eq. \ref{ecthrel}, $n_{F}$ represents the Fermi-Dirac distribution and $\mu$ is the chemical potential. This above relationship, which holds in a magnetic field, was also shown to be valid in the presence of disorder in Ref. \onlinecite{girvinjonsonprb}, where both the off-diagonal and diagonal conductivity on an IQH plateau were calculated treating disorder within SCBA. The longitudinal electrical conductivity is
\begin{eqnarray}
\label{econd}
L^{11}_{rr} &=& \frac{e^{2}\omega_{c}^2}{8 \pi h} \int_{-\infty}^{\infty} d \epsilon \bigg( -\frac{\partial n_{F}}{\partial \epsilon} \bigg)\\
\nonumber
& &  \sum_{n}  [ (n+1)Im G_{n} (\epsilon) Im G_{n+1} (\epsilon) ],
\end{eqnarray}
where $n$ corresponds to the Landau level (LL) index, $\omega_{c}=eB/m $ is the cyclotron frequency and $G_{n}(\epsilon)$ is the SCBA dressed Green's function for the $n^{th}$-LL given by
\begin{equation}
G_{n} (z) = \frac{1}{z - \hbar \omega_{n} -\Sigma(z)},
\end{equation}
where $\hbar \omega_{n} = (n+1/2) \hbar \omega_{c}$. Assuming a random impurity potential $V$ with a white noise distribution $\langle V ({\bf r}) V({\bf r}') \rangle_{av} = 2 \pi l_{B}^2 \hbar^2 \sigma^{2} \delta ({\bf r}-{\bf r}')$ the SCBA self-energy $\Sigma $ in the high field approximation ($\sigma << \omega_{c}$) can be written as
\begin{equation}
\Sigma (\omega + i \delta) = \frac{\omega - \omega_{L}}{2} - i \sigma \sqrt{1- \bigg(\frac{\omega - \omega_{L}}{2 \sigma}\bigg)^2},
\end{equation}
where $\omega_{L} $ is the energy of the LL nearest to $\epsilon$ and $\sigma$ is a measure of the white noise disorder potential. The disorder broadened density of states (DOS) is given as
\begin{equation}
\rho_{n} (\epsilon) = \frac{1}{2 \pi^2 l_{B}^2 \hbar \sigma} \sqrt{1- \bigg(\frac{\epsilon - \omega_{L}}{2 \hbar \sigma} \bigg)^2},
\end{equation}
is semi-elliptical with a semi-minor axis $2 \hbar \sigma$. The disorder broadened LL density of states for the lowest LL is plotted in Fig. \ref{figthree} which also indicates the quasiparticle/quasihole regimes.\\
\indent
The expression (\ref{econd}) can be used in (\ref{ecthrel}) to calculate the longitudinal thermal response $L^{12}_{rr}$ which can then be used to calculate the Corbino thermopower given by (\ref{Cth}).
For an isolated LL, it can be expressed as a two-paramter scaling function of $(\mu - (n+1/2)\hbar \omega_{c})/(\hbar \sigma) $ and $\hbar \sigma /k_{\beta} T$ with the scaling function plotted for different values of $\hbar \sigma /k_{\beta} T$ in Fig. \ref{figfour}. In order to better understand features the Corbino thermopower of an isolated LL it is best to orient oneself and understand the disorder broadened density of states (DOS) which in the SCBA is  semi-elliptical with a semi-minor axis $2 \hbar \sigma$ as indicated in Fig. \ref{figthree}. Away from the tails of the disorder broadened DOS the thermopower indicates divergent behavior as a function of the temperature $T$, reminiscent of an insulator. This behavior is anticipated away from the tails of the disorder broadened DOS and near the center of the QH plateau where quasielectrons (quasiholes) are thermally activated and the thermopower should naively correspond to the entropy per quasiparticle per quasiparticle charge.\\
\indent
Furthermore, the Corbino themopower changes sign at the position of the LL where the disorder broadened DOS is maximized. This sign change is due to the fact that transport switches from being electron-dominated to hole-dominated regimes, as a result the charge of the relevant quasiparticles change sign. The overall features of the Corbino thermopower of an isolated LL already indicates behavior anticipated by the {\it ``entropy per quasiparticle per quasiparticle charge"} definition.\\
\indent
We plot in Fig. \ref{figfive} Corbino thermopower with contributions from all LLs taken into account, as a function of the chemical potential for fixed disorder strength $\hbar \sigma =0.05\hbar\omega_c$ at different temperatures. As expected, we see a periodic behavior from one LL to another, further justifying the absence of the QH condensate contribution.  The sign change in the thermopower when the chemical potential in half-way between two neighboring LLs, which corresponds to the {\em center} of a QH plateau, has a similar origin to the sign change at the QH transition points discussed above:
This is due to the fact that transport to the right of the center of the QH plateau is dominated by thermally activated electrons in the higher LL, and holes to the left in the lower LL. As mentioned earlier in this regime the quasiparticle contribution to the transport coefficients exhibit Hall insulating behavior and as we show in the next section the thermopower for low temperatures is equal to the {\it ``entropy per quasiparticle per quasiparticle charge"}.  In the next section we establish that this relation is exact at low temperatures near the center of the an integer QH plateau. \\
\indent
\subsection{Corbino thermopower and entropy per quasiparticle}
Next we analyze the entropy per quasiparticle per quasiparticle charge and its relation to the Corbino thermopower. The entropy is the partial derivative of the grand canonical potential $\Omega $ with respect to the temperature $T$, $S = -\partial \Omega /\partial T$. The grand canonical potential in the presence of the disorder broadened LL DOS for an isolated LL can be written as
\begin{equation}
\label{thermodynamicpot}
\Omega = -k_{B}T \frac{N_{\phi}}{\pi \sigma } \sum_{n} \int_{-2 \hbar \sigma + \hbar \omega_{n} }^{2 \hbar \sigma + \hbar \omega_{n}} d \epsilon \log(1 + e^{\beta(\mu - \epsilon)} ) \rho_{n}(\epsilon),
\end{equation}
where $N_{\phi} = A/2 \pi l_{B}^2$ is the degeneracy of a single LL. In the following we concentrate on a single LL only, neglecting the effects of LL mixing. This assumption can be formally justified by working the in the high field approximation $\hbar \omega_{c} >> \hbar \sigma$. In this limit the entropy is a periodic function of the chemical potential $\mu$.
The total entropy for an isolated LL $ S = S^{+} + S^{-}$ can be split into contributions from quasielectrons and quasiholes $S^{\alpha}$, $\alpha = +(-)$ for quasielectrons (quasiholes). It is advantageous for what follows to  express $S^{\alpha}$ in terms of the number of quasielectrons (quasiholes) $N_{q}^{\alpha}$ as
\begin{equation}
\label{altent}
S^{\alpha}(T,\mu) = (-1)^{\alpha +1} \int_{-2 \sigma \alpha + \hbar \omega_{c}/2}^{\hbar \omega_{c} /2} d\epsilon  \frac{(\epsilon - \mu)}{k_{\beta} T^2} \frac{\partial n_{F}}{\partial \epsilon} N_{q}^{\alpha} (\epsilon),
\end{equation}
where we have used the fact that $\rho_{n}(\epsilon) = \partial N_{q}(\epsilon)/\partial \epsilon$ and performed an integration by parts on Eq. (\ref{thermodynamicpot}), and then taken the derivative with respect to the temperature. The number of quasiparticles $N_{q}^{\alpha}$ can be expressed as
\begin{equation}
N_{q}^{\alpha}(\mu) = (-1)^{\alpha}\frac{N_{\phi}}{\pi \sigma} \int_{-2 \hbar \sigma \alpha + \hbar \omega_{c}/2}^{\mu_{\alpha}} d\epsilon \sqrt{1-\bigg( \frac{\epsilon-1/2 \hbar \omega_{c}}{2 \hbar \sigma}\bigg)^2}
\end{equation}
with $\mu_{+} \leq \hbar \omega_{c}/2$ (electron dominating) and $\mu_{-} \geq \hbar \omega_{c}$ (hole dominating), and with our definition we always have $N_{q}^{\alpha} > 0$. The thermally activated quasiparticle can then be expressed as
\begin{equation}
\label{thnquasiparticle}
N_{q}^{\alpha}(T,\mu) =  \int_{-\infty}^{\infty} d\epsilon \frac{\partial n_{F}}{\partial \epsilon} N_{q}^{\alpha}(\mu).
\end{equation}
\indent
To compare the Corbino thermopower calculated in the previous section to the {\it ``entropy per quasiparticle per quasiparticle charge"} in the low temperature limit we start by identifying two different regions as shown in Fig \ref{figthree}. For the sake of clarity we now restrict our analysis to quasielectrons only, since the analysis for quasiholes is similar. We relax this constraint in our final answers which are general and apply equally to both types of quasiparticles.  Regions I and II for quasielectrons are given as $0<\mu<-2\sigma+ \hbar \omega_{c}/2$ and $-2\sigma+ \hbar \omega_{c}/2 < \mu < \hbar \omega_{c}/2$ respectively(see Fig. ~\ref{figthree}). In region I the quasiparticles are purely thermally activated and obey Boltzmann statistics, the diagonal response functions exhibit behavior similar to an insulator and hence divergent thermpower (see Fig. ~\ref{figfour}). In contrast, region II contains a finite density of quasielectrons even at $T=0$ and the thermopower depends in particular on the details of disorder strength and the chemical potential. \\
\indent
For region I working in the low temperature limit $k_{\beta} T \ll -2 \sigma- \mu + \hbar \omega_{c}/2 $ the quasiparticles obey Boltzmann statistics and one can replace the derivative of the Fermi function $\partial n_{F}/\partial \epsilon $ in Eqs. (\ref{ecthrel}), (\ref{econd}) and (\ref{altent}) by $e^{\beta(\epsilon - \mu)}$.
In this limit the entropy can be expressed as (a similar situation arises for quasiholes),
\begin{equation}
S^{+} = - \frac{\partial}{\partial T} \bigg[ e^{\beta(\mu +2 \sigma -1/2)} \int_{0}^{\delta} d \epsilon e^{-\beta \epsilon } N^{+}_{q} (T=0,\epsilon \to 0 ) \bigg]
\end{equation}
where $\delta $ is a small region where the derivative of the Fermi function overlaps with the LL broadened density of states and with leading order behavior for $N^{+}_{q}(T=0,\epsilon \to 0) = 2/3(x/\sigma)^{3/2}$. With the same approximation for Eq. \ref{thnquasiparticle}
the {\it ``entropy per quasiparticle per quasiparticle charge"} in Region I can be expressed as,
\begin{eqnarray}
\nonumber
\frac{S^{\alpha}}{e N_{q}^{\alpha}} &=& (-1)^{\alpha} \frac{k_{\beta}^2 T}{e}\frac{\partial}{\partial T}\bigg[ \log \bigg( e^{\beta(\mu +2 \sigma \alpha -1/2)} \\
& & \int_{0}^{\delta/(k_{\beta} T)} d u e^{- u} u^{3/2}\bigg) \bigg].
\end{eqnarray}
\begin{figure}[t]
\begin{center}
\includegraphics[width=3.2in,height=2.3in]{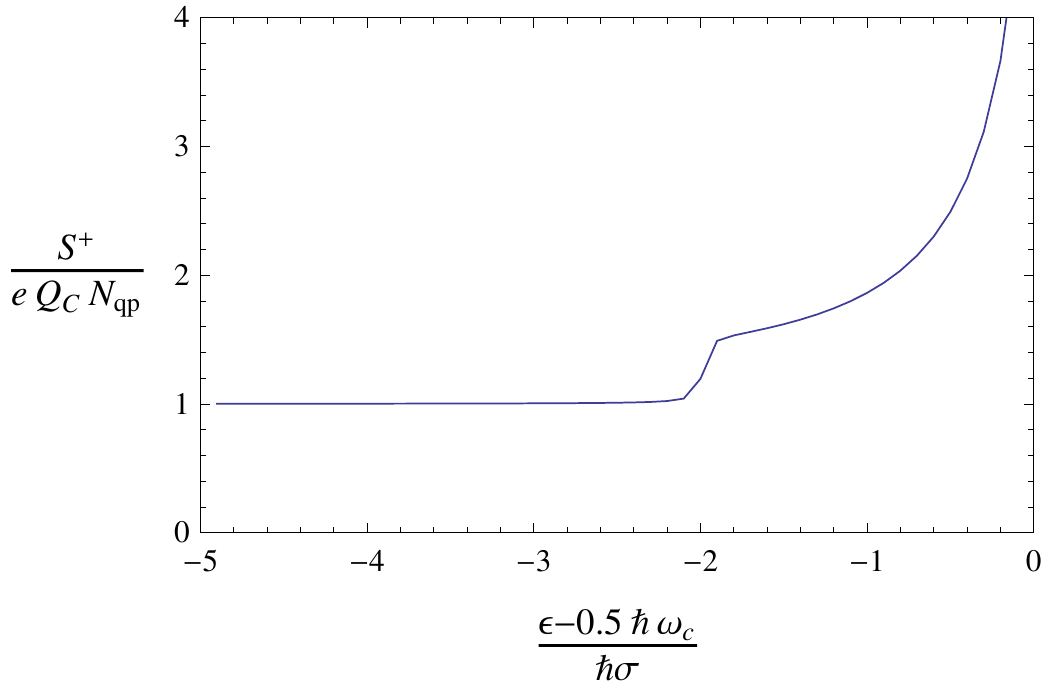}
\caption{Numerical attained ratio of the {\it entropy per quasiparticle per quasiparticle charge} $S/(e N_{q})$  and the Corbino thermopower $Q_{C}$ for the quasi-electron region of an isolated lowest Landau level for $k_{\beta} T/\hbar \sigma = 0.01$.}
\label{figsix}
\end{center}
\end{figure}
Similarly one can use the same approximation to evaluate Eqs. (\ref{ecthrel}) and (\ref{econd}) in Region I giving the Corbino thermopower near the center of the QH plateau,
\begin{eqnarray}
\nonumber
Q_{C}^{\alpha} &=& (-1)^{\alpha} \frac{k_{\beta}^2 T}{e}\frac{\partial}{\partial T}\bigg[ \log \bigg( e^{\beta(\mu +2 \sigma \alpha -1/2)} \\
& & \int_{0}^{\delta/(k_{\beta} T)} d u e^{- u} u \bigg) \bigg]
\end{eqnarray}
where $\alpha =+(-)$ gives the Corbino thermopower expression to the right (left) of the center of the QH plateau in the low temperature regime. This gives an explicit verification of the interpretation of the Corbino thermpower as the {\it ``entropy per quasiparticle per quasiparticle charge" }
\begin{equation}
Q_{C}^{\alpha}  =\frac{S^{\alpha}}{e N_{q}^{\alpha}} = \alpha \frac{k_{\beta}^2}{e} \frac{(\mu + 2\sigma \alpha -1/2)}{T} + \cdots.
\end{equation}
\indent
In the opposite regime (region II) where the quasiparticles resemble a degenerate Fermi gas for $k_{B}T \ll \hbar \sigma$, we recover a Mott like expression for the Corbino thermopower,
\begin{equation}
Q_{C} = \alpha \frac{k_{\beta}^2}{e} T \frac{\pi^2}{3} \frac{1}{L^{11}_{xx}}\frac{dL^{11}_{xx}}{d \epsilon}\bigg{|}_{\epsilon = \mu} .
\end{equation}
Similarly the {\it entropy per quasiparticle per quasiparticle charge} be expressed as
\begin{equation}
\frac{S^{+}}{eN^{+}} = \alpha \frac{k_{\beta}^2}{e} T \frac{\pi^2}{3} \frac{1}{N^{+}_{q}}\frac{dN^{+}_{q}}{d \epsilon}\bigg{|}_{\epsilon = \mu}.
\end{equation}
In the latter case while we do not have an exact equality, we still find $Q_C$ scales like the {\it entropy per quasiparticle per quasiparticle charge} with the constant of proportionality depending on the specific model for disorder, and other details. The deviation from equality gets progressively worse as the chemical potential approaches the LL energies, where the distinction between quasielectron and quasihole becomes ambiguous. This is evident in Fig. \ref{figsix} where the numerically attained ratio of {\it entropy per quasiparticle per quasiparticle charge} $S/(eN_{q})$ and Corbino thermopower $Q_C$ is plotted as a function of the chemical potential $\mu$. Fig. \ref{figsix} also indicates the two are equal in Region I which is near the center of the QH plateau. In Region II the two are no longer equal; however they have the same (linear) $T$ dependence and hence a constant ratio in the low $T$ limit.\\

\section{Application to non-Abelian Quantum Hall States}

It is clear that Eq. (\ref{eq:centralresult}) can be used to probe entropy carried by non-Abelian quasiparticles in non-Abelian QH liquids, which is dominated by the topological entropy $S_D$ at low temperature. The best place to do this is near the center of the QH plateau, where the quasiparticle density is low. This is {\em opposite} to the case of Hall bar geometry,\cite{yanghalperin} in which case thermopower measures {\em total} entropy carried by the quasiparticles (divided by number of {\em electrons}), one thus needs to be near the edge of QH plateau to have higher quasiparticle density and thus entropy. Here since it is entropy {\em per quasiparticle} that is measured, the low-density regime is preferable. There are several advantages of working near the plateau center, where the physics in general is simpler. Among them, we mention: (i) We do not need to worry about the residue coupling among quasiparticles that can lift the ground state degeneracy, as they decay exponentially with quasiparticle separation. (ii) We do not need to worry about competing states that may appear (possibly in parts of the sample), which can carry substantial entropy.\\
\indent
Quantitatively, we expect Corbino thermopower to saturate to a {\em finite} value in the low temperature limit that depends on the quasiparticle's quantum dimension:
\begin{equation}
Q_C(T\rightarrow 0)=(k_\beta/e^{\star})\log d.
\label{eq:nonabelianthermopower}
\end{equation}
For the specific case of $5/2$, we expect $|e^{\star}|=|e|/4$ and $d=\sqrt{2}$,\cite{mooreread} thus
\begin{equation}
|Q_C(T\rightarrow 0)|=(4k_\beta/e)\log\sqrt{2}\approx 1.2\times 10^{-4} V/K,
\end{equation}
which is at least one order of magnitude larger than what has been observed in the FQH regime below $0.1K$. In the Corbino geometry, the sign of $Q_C$ changes when one moves through the plateau center, as one goes from the quasiparticle dominated to quasihole dominated regime.
To approach the saturation value of Eq. (\ref{eq:nonabelianthermopower}), we need the temperature to be sufficiently low such that $S_n(T)\ll S_D$. Assuming that the quasiparticles form a Wigner crystal in a completely clean sample, this happens for $T\ll T_D$, where the characteristic temperature $T_D$ is the Debye temperature of the Wigner crystal, estimated in Ref. \onlinecite{yanghalperin} [see its Eq. (13)]. We note this would be a {\em lower bound} for realistic samples, as disorder can pin the Wigner crystal (or perhaps individual quasiparticles), and open gaps in the magnetophonon spectra; this tends to {\em suppress} $S_n(T)$ at low temperature, and pushes the saturation temperature higher.

One caveat to keep in mind is for samples with some inhomogeneity, there are preexisting quasiparticles and quasiholes (with equal average density) at the center of the QH plateau, and their contributions to $Q_C$ would cancel due to their opposite charge. In that case one does need to move away from the plateau center, such that the number of quasiparticles or quasiholes induced by the deviation dominates the preexisting ones. This may put some stringent constraints on the sample quality for the observability of the behavior indicated in Eq. (\ref{eq:nonabelianthermopower}).

\acknowledgements
One of us (KY) thanks R. R. Du for a useful correspondence that motivated the present work.
This work was supported in part by NSF grant DMR-1004545 (KY).

\end{document}